\documentclass[prb,aps,floats,superscriptaddress]{revtex4}
\usepackage{amsmath}
\usepackage{graphicx}

\newcommand{\mr}[1]{{{\mathrm{#1}}}}
\newcommand{\mcal}[1]{{\mathcal{#1}}}
\newcommand{\dt}{\partial_\tau}
\newcommand{\inte}{\int_0^\beta \!\!\!\! \mr{d}\tau}
\newcommand{\w}{\omega}
\newcommand{\wn}{\omega_n}
\newcommand{\s}{{\sigma}}
\newcommand{\zl}{Z_\lambda}
\newcommand{\gl}{g_\lambda}

\begin{document}
\draft

\title{Non-Fermi liquid behavior from two-dimensional antiferromagnetic
fluctuations: a renormalization-group and large-N analysis}
\author{Sergey Pankov}
\affiliation{Center for Materials Theory, Department of Physics and Astronomy,\\
Rutgers University, Piscataway, New Jersey 08854, USA}
\author{Serge Florens}
\affiliation{Laboratoire de Physique Th\'eorique, Ecole Normale Sup\'erieure,\\
24 Rue Lhomond, 75231 Paris Cedex 05, France}
\author{Antoine Georges}
\affiliation{Laboratoire de Physique Th\'eorique, Ecole Normale Sup\'erieure,\\
24 Rue Lhomond, 75231 Paris Cedex 05, France}
\author{Gabriel Kotliar}
\affiliation{Center for Materials Theory, Department of Physics and Astronomy,\\
Rutgers University, Piscataway, New Jersey 08854, USA}
\author{Subir Sachdev}
\affiliation{Department of Physics, Yale University,\\
P.O. Box 208120, New Haven CT 06520-8120, USA}
\date{\today}

\begin{abstract}
We analyze the Hertz-Moriya-Millis theory of an antiferromagnetic quantum
critical point, in the marginal case of two dimensions $(d=2,z=2)$. Up to
next-to-leading order in the number of components ($N$) of the field, we
find that logarithmic corrections do not lead to an enhancement of the Landau
damping. This is in agreement with a renormalization-group analysis,
for arbitrary $N$. Hence, the logarithmic effects are unable to account for the behavior
reportedly observed in inelastic neutron scattering experiments on
${\rm CeCu}_{6-x}{\rm Au}_x$. We also examine the extended dynamical mean-field
treatment (local approximation) of this theory, and
find that only subdominant corrections to the Landau
damping are obtained within this approximation, in contrast to recent claims.
\end{abstract}

\maketitle

\section{Introduction}
Materials in the vicinity of a quantum critical
point (QCP) continue to be the subject of intensive investigations \cite{StewartRMP}.
Particular attention has been paid to the
heavy fermion system ${\rm CeCu}_{6-x}{\rm Au}_x$
which undergoes a phase transition from a paramagnetic, heavy-fermion metal
to an antiferromagnetic metal as a function of chemical composition and pressure
\cite{HvL1994,Schroder2000}.
Another remarkable case is the stoichiometric compound ${\rm YbRh_2Si_2}$ which is
apparently located very close (on the magnetic side) of an antiferromagnetic QCP
\cite{Trovarelli2000}. Hydrostatic pressure stabilizes the magnetic phase in that case.
As the critical point is approached, these systems
exhibit an  enhancement of the specific heat coefficient $C/T$,
magnetic susceptibility, resistivity $\Delta\rho$ and thermoelectric power, which is not easily
described within the conventional theory of a three dimensional quantum
critical antiferromagnet.

Neutron scattering experiments\cite{Rosch,Schroder1998,Stockert1998,Schroder2000}
on ${\rm CeCu}_{6-x}{\rm Au}_x$ have revealed that the spin fluctuation spectrum of
this compound has a {\it two-dimensional} character in a wide range of temperatures.
Indeed, maxima of the neutron scattering intensity are observed along
rod-like structures in reciprocal space along which spin
fluctuations are independent of one component of the wavevector.
This motivated Rosch et al.\cite{Rosch} to propose that the observed anomalies
are the results of the coupling of three-dimensional electrons to
quasi two-dimensional critical spin fluctuations. In such a situation, conventional
Hertz-Moriya-Millis theory\cite{hertz,moriya,millis,sachdev}
of quantum critical behavior leads to
$C/T\propto \ln(1/T)$, $\Delta\rho\propto T$, consistent with experimental
findings. A similar observation applies to the thermoelectric
power \cite{ipaul}. The microscopic reason for the two-dimensional nature of spin
fluctuations is not clear however for ${\rm CeCu}_{6-x}{\rm Au}_x$, while magnetic frustration
might provide a natural explanation\cite{Trovarelli2000} in the case of ${\rm YbRh_2Si_2}$ (for which
inelastic neutron scattering data are not yet available).

The approach of Rosch et al.\cite{Rosch} requires fine tuning on a microscopic
level (since a three- dimensional dispersion of the electrons will generically
lead to three- dimensional critical fluctuations at the QCP) but it is
internally consistent on a theoretical level.
Indeed, the vertex corrections to the electron- spin fluctuation interaction
is finite \cite{ipaul}
in the case of a coupling to a three-dimensional fermion spectrum.
(Note that the situation is different for two-dimensional fermions, which
leads to a singular vertex\cite{chubukov}).
Since the vertex is non-singular,
the physics of long-wavelength magnetic fluctuations is
described by a $\phi^4 $ field theory  with ohmic damping.
The quartic term is marginally irrelevant for $d=2$, which is the upper critical
dimension of this theory. Hence, the flow of the quartic term will lead to
{\it logarithmic deviations} from mean-field critical behavior.
These effects have not been investigated previously for the frequency-dependent
response function in this context.
Inelastic neutron scattering \cite{Schroder1998,Stockert1998,Schroder2000}
on ${\rm CeCu}_{6-x}{\rm Au}_x$ at the QCP ($x_c\simeq 0.1$)
have revealed an anomalous enhancement of the Landau damping.
Whether logarithmic effects at the upper critical dimension could account for such
an enhancement in the range of experimentally accessible frequencies
is an outstanding open question which we address in this paper.
We use the number of components ($N$) of the field as a control parameter, and
perform a calculation of the dynamical susceptibility to order $1/N$.
We find that, to this order, the sign of the corrections to Landau damping
{\it cannot explain the experimental data}.

An alternative theoretical viewpoint on the physics at the QCP is
that the spin-fluctuation self-energy at the QCP is purely local
(i.e momentum independent) and has an anomalous power-law
dependence on frequency and temperature. This was first
proposed\cite{Schroder1998,Schroder2000} as a phenomenological fit
to the inelastic neutron scattering data in the form: ${\rm
Im}\chi^{-1}_{q=Q} = \omega^\alpha f(\omega/T)$ with $\alpha\simeq
0.7-0.8$. Si et al.\cite{SiNature} have further developed this
point of view in the framework of the extended dynamical
mean-field (EDMFT) theory of the Kondo
lattice\cite{sismith,kaju,sengupta}. It has been argued in this
context that the separation of electronic degrees of freedom and
long-wavelength magnetic fluctuations is not
legitimate\cite{SiNature,coleman_poland}, because the (Kondo)
coherence scale below which electron degrees of freedom can be
eliminated might vanish at the QCP. Recently however, Grempel and
Si showed\cite{GrempelSi} that {\it starting above the Kondo
temperature} the EDMFT of the Kondo lattice can be mapped, using
bosonization tricks, onto the EDMFT of the $\phi^4$ theory for
magnetic modes (see also the earlier work by Sengupta and one of
us\cite{sengupta}). They further suggested that, in two
dimensions, this theory would lead to an anomalous Landau damping,
with the power-law form mentioned above ($\alpha<1$). We study
this effective theory in the EDMFT approximation, using both a
strict 1/N expansion and a numerical solution using the large-N
expansion as an ''impurity solver''. We find that, contrary to the
statements of Ref.~\onlinecite{GrempelSi}, {\it no anomalous
power-law is generated}. In fact, the EDMFT approximation to the
$\phi^4$ field-theory in the marginal case does not correctly
reproduce the logarithmic corrections to the damping rate found in
our direct calculation for $d=2$, and leads only to subdominant
$O(\omega^2)$ corrections at low-energy. This extends to the
marginal case the previous work of Motome and two of
us\cite{pankov} on the the EDMFT approximation applied to critical
behavior, in which it was indeed found that the approximation is
better in higher dimensions, and becomes unreliable below the
upper critical dimensions.

Hence, the general conclusion of our paper is that logarithmic effects at the
upper critical dimensions in the conventional theory of an antiferromagnetic QCP cannot
explain the behavior observed experimentally for ${\rm CeCu}_{6-x}{\rm Au}_x$, at
least at dominant order in the $1/N$ expansion.

This paper is organized as follows: in section~\ref{sec_rg}, we
perform a general renormalization-group analysis of the $\phi^4$
theory with damping. In section~\ref{largeNlimit}, we take a
large-$N$ perspective, and present an
explicit solution up to next to leading order in the 1/N
expansion. We also describe the local (EDMFT) approximation to this theory, up to
order $1/N$.
In section~\ref{fullN} we discuss in more details the EDMFT approach,
and solve numerically the EDMFT equations using a large-N impurity solver.
We conclude with a discussion of the
implications of our work for the understanding of non-Fermi liquid behavior
in heavy-fermion materials close to an antiferromagnetic QCP.

\section{Renormalization group analysis}
\label{sec_rg}

\subsection{Model and general renormalization-group framework}
\label{sec:model}

In this paper, we consider the following field theory\cite{hertz,millis,sachdev}
for an $N$-component field $\phi_a$ ($a=1,\cdots,N$):
\begin{equation}
\label{action}
S={\frac{1}{2}}\sum_{q,\omega}\,D^{-1}_0(q,i\omega)\,\sum_a
\phi_a^2(q,i\omega)\, +\,{\frac{u_0}{4!}}\,\int d^dx
d\tau\,[\sum_a \phi_a(x,\tau)^2]^{\,2}
\end{equation}
in which the free propagator reads, for imaginary frequencies:
$D^{-1}_{0}(q,i\omega)=r+|\omega|+q^2$. This theory describes the
vicinity of an antiferromagnetic quantum critical point,
corresponding to the (bare) value $z=2$ of the dynamical critical
exponent. In this context, the dissipative term is induced by the
coupling to electronic degrees of freedom (Landau damping). The
$q$-vector is the difference from the ordering wavevector, and $r$
is a parameter which measures the distance to the critical point.
We are particularly interested in this paper in the
two-dimensional case, but we shall also briefly consider lower
dimensions $1\leq d<2$ (Sec.~\ref{sec:rg_epsilon}). In this
section, we shall use the following normalized coupling constant:
\begin{equation}
g_0 \equiv S_d u_0\,\,\, {\rm with:}\,\,\,
S_d=\frac{2}{\pi\Gamma(d/2)(4\pi)^{d/2}} \label{coupling_rg}
\end{equation}
When considering the bare theory, an (ultra-violet) cutoff
$\Lambda$ on momenta is introduced. We consider the bare
irreducible 2-point function $\Gamma\equiv \chi^{-1}$, which is
the inverse of the correlation function (dynamical susceptibility
$\chi(q,i\omega)=\langle \phi(q,i\omega)\phi(q,-i\omega)\rangle$).
The renormalization group (RG) specifies\cite{zinn} how $\Gamma$
changes upon a rescaling of the cutoff $\Lambda\rightarrow
\lambda\Lambda$. The corresponding RG equation is obtained by
imposing that the renormalized 2-point function $\Gamma_R= Z
\Gamma$ is independent of the original cutoff (with $Z$ the field-
or ''wave function'' renormalization). Focusing first, for
simplicity, on the field-theory (\ref{action}) in the critical
(massless) case, $\Gamma$ satisfies the following equation (all
Green's functions in this and the next subsection are at $T=0$):
\begin{equation}
\Gamma\left[q,i\omega;g_0,\Lambda\right] = (\lambda\Lambda)^2
Z_\lambda^{-1}\, \Gamma\left[\frac{q}{\lambda\Lambda},
\frac{Z_\lambda}{(\lambda\Lambda)^2}\,i\omega;g_\lambda,\Lambda=1\right]
\label{eq:rg}
\end{equation}
In this expression, $Z_\lambda$ and the running coupling constant $g_\lambda$ are
defined from the usual RG functions:
\begin{equation}
\lambda\frac{d}{d\lambda}\,\gl = \beta(\gl)\,\,\,,\,\,\,g_{\lambda=1}=g_0
\label{def:gl}
\end{equation}
\begin{equation}
\lambda\frac{d}{d\lambda}\,\ln \zl = \eta(\gl)\,\,\,,\,\,\,Z_{\lambda=1}=1
\label{def:zl}
\end{equation}
Let us emphasize a key aspect of Eq.(\ref{eq:rg}), namely that the
damping term does not require an independent renormalization so
that the frequency dependence in the r.h.s of (\ref{eq:rg})
involves only the wave-function renormalization $\zl$. This can be
proven to hold to all orders, using a field theoretic
RG\cite{aharony}. For a use of this method in a different context
see Ref.\onlinecite{boyanovsky}

\begin{figure}[htbp]
\begin{center}
\vspace{0.6cm}
\includegraphics[width=10cm,angle=0]{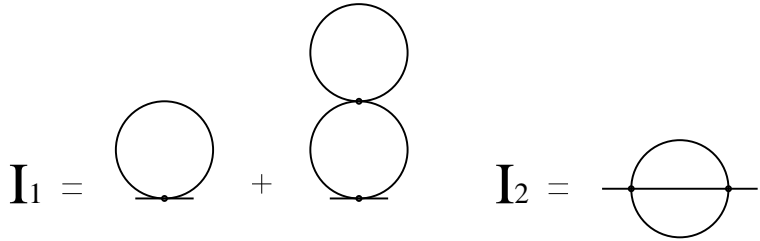}
\end{center}
\caption{Graphs contributing to the 2-point function at two-loop order.}
\label{I1I2}
\end{figure}

We have calculated the RG function $\beta$ at one-loop order and the function
$\eta$ at two-loop order. The diagrams contributing to the 2-point
function $\Gamma$ up to this order are depicted in Fig.~\ref{I1I2}.
We obtain (with $\epsilon=d-2$):
\begin{eqnarray}
&&\beta(g)=-\epsilon g+b_2 g^2+O(g^3)
\nonumber\\
&&\eta(g)=c_2 g^2+O(g^3)
\nonumber
\label{beta_eta}
\end{eqnarray}
where the coefficients $b_2$ and $c_2$ read, in the N-component theory:
\begin{equation}
b_2={\frac{N+8}{6}}\,\,\,\,,\,\,\,c_2=\frac{(N+2)(12-\pi^2)}{144}
\label{b2c2}
\end{equation}
Details of calculation of the coefficient $c_2$ are given in
Appendix \ref{rgflowappx}.

\subsection{The marginal case $d=2$: logarithmic corrections to Landau damping}
\label{sec:rg_marginal}

We focus here on the marginal case $d=2$ at the quantum-critical
point, and integrate the above RG equations in order to obtain the
corrections to the Landau damping at $T=0$. Integrating the RG
equations (\ref{def:gl},\ref{def:zl}) using (\ref{beta_eta})
yields, to this order:
\begin{eqnarray}
&&g_\lambda={\frac{g_0}{1-b_2 g_0 \ln{\lambda}}}
\nonumber\\
&&Z_\lambda=\exp\left({\frac{c_2 g_0^2 \ln{\lambda}}{1-b_2 g_0 \ln{\lambda}}}\right)
\label{gZ}
\end{eqnarray}
As expected in the marginal case, $\gl$ flows logarithmically to zero
as $\lambda\rightarrow 0$, while $\zl$ tends to some (non-universal)
constant $Z^*$.
To analyze the frequency dependence of $\Gamma$ we use the
general RG equation (\ref{eq:rg}), setting $q=0$ and choosing
$\lambda=\lambda^*$ such that
$\omega Z_{\lambda^*}/(\lambda^*\Lambda)^2=1$. This leads to:
\begin{equation}
\Gamma(q=0,i\omega;g_0,\Lambda)= |\omega|\,\Psi(g_{\lambda^*})
\label{freqdep}
\end{equation}
in which we use the notation
$\Psi(g)\equiv\Gamma(q=0,i\omega=i;g,\Lambda=1)$. We then expand
the r.h.s of (\ref{freqdep}) in powers of $g_{\lambda^*}$, with:
$g_{\lambda^*}\sim -1/(b_2\ln\lambda^*)\sim
-2/[b_2\ln(Z^*|\omega|/\Lambda^2)]$. This expansion actually
starts at {\it second-order} in $\gl$, since the 2-point function
$\Gamma$ is subtracted to insure that one sits at the critical
point (in other words, the tadpoles $I_1$ do not contribute to the
frequency dependence). Noting that:
$I_2(q=0,i\omega=0;\gl,\Lambda)- I_2(q=0,i\omega,\gl,\Lambda) =
c\gl^2|\omega|$, coefficient $c$ computed in Appendix
\ref{rgflowappx},
 we finally obtain a correction to the frequency
dependence of the 2-point function at the $T=0$ QCP of the form:
\begin{equation}
\Gamma(q=0,i\omega) = |\omega| \left( 1 +
\frac{\left[6\pi^2\ln{2}-11\zeta(3)\right](N+2)} {12\,(N+8)^2}\,
\frac{1}{\ln^2(|\omega|Z^*/\Lambda^2)} \right)
\label{rg_correction}
\end{equation}
Hence, only subdominant corrections to the Landau damping are
generated in the marginal case $d=2$. Furthermore, the positive
coefficient in the above expression can be interpreted as an
effective exponent $\alpha=2/z>1$. This is in agreement with
the $1/N$ expansion, which we consider later in this paper.

\subsection{$\epsilon=2-d$ expansion of critical exponents and corrections to scaling}
\label{sec:rg_epsilon}

We consider here the case $d<2$, in which the quartic term is
relevant, and the effective coupling $\gl$ tends to a non-trivial
fixed point. The case $d=1$ is relevant, for example, when
considering\cite{sachdev} an Ising chain in a transverse field
($N=1$) or a chain of quantum rotors ($N>1$) coupled to a
dissipative environment.

We make use of the general RG equation (\ref{eq:rg}) by choosing
$\lambda=\lambda^*$ such that $q/\lambda^*\Lambda =1$.
For $\lambda\rightarrow 0$, we now have: $\gl\rightarrow g^*$ and
$\zl\sim \lambda^\eta$ (with the critical exponent $\eta\equiv \eta(g^*)$.
Hence, we obtain (at $T=0$) for the dynamical susceptibility
$\chi=\Gamma^{-1}$ at low-frequency and small momentum:
\begin{equation}
\chi(q,\omega) = q^{-2+\eta}\phi(\omega/cq^{2-\eta})
\end{equation}
In this expression, $\phi$ is a universal scaling function associated with the
fixed point. From this expression, the dynamical critical exponent is identified as:
\begin{equation}
z=2-\eta
\end{equation}
An expression which holds to all orders, and is the result of the existence of a unique
RG function, as explained above.
This scaling form can actually be generalized to finite temperature in the quantum critical
regime in the form:
\begin{equation}
\chi(q,\omega) = \frac{1}{T}\,\Phi\left(\frac{\omega}{T},\frac{c_1 q}{T^{1/z}}\right)
\end{equation}
with $\Phi$ a universal scaling function, and $c_1$ a
non-universal constant. In particular, at zero-momentum,
$T\chi(q=0,\omega)$ is an entirely universal scaling function of
$\omega/T$. At small $\omega/T$, we expect an analytic dependence
on $\omega/T$, with $\Gamma (0, \omega)/T = \mathcal{C}_1 - i
\mathcal{C}_2 \omega/T + \ldots$. In contrast, at large $\omega/T$
we have $\Gamma (0, \omega)/T = -i\mathcal{C}_3 \omega/T$, and the
subdominant terms are not analytic, but are related to those in
(\ref{subdominant}) below. Here $\mathcal{C}_{1-3}$ are all
non-trivial {\em universal} constants. The value of
$\mathcal{C}_3$ is determined in the $\epsilon$ expansion by the
computations in Appendix~\ref{rgflowappx} ($\mathcal{C}_3 = 1 + c
(\epsilon/b_2)^2 + \ldots$), and this universality is clearly
related to the universal logarithmic correction in
(\ref{rg_correction}). The values of $\mathcal{C}_{1,2}$ require a
computation of the damping at $T>0$, and accurate determination of
these is likely to require a self-consistent treatment of the loop
corrections, as discussed in Ref~\onlinecite{sachdevrg}. In
particular, $\mathcal{C}_2 \neq \mathcal{C}_3$ and hence
\begin{equation}
\lim_{\omega\rightarrow 0} \lim_{T \rightarrow 0} \frac{\mbox{Im}
\Gamma (q,\omega)}{\omega} \neq \lim_{T\rightarrow 0} \lim_{\omega
\rightarrow 0} \frac{\mbox{Im} \Gamma (q,\omega)}{\omega}.
\label{noncommute}
\end{equation}
So the Landau damping co-efficient is sensitive to the order of
limits of $\omega \rightarrow 0$ and $T \rightarrow 0$; we expect
that similar considerations will also apply to the logarithmic
corrections in $d=2$ noted in (\ref{rg_correction}).

From the RG expressions in Section~\ref{sec:model}, the
$\epsilon$-expansion of the critical exponent $\eta=2-z$ reads, to
order $\epsilon^2$:
\begin{equation}
\eta = \frac{(N+2)(12-\pi^2)}{4(N+8)^2}\, \epsilon^2 + O(\epsilon^3)
\end{equation}
We have also obtained the correlation length exponent at order $\epsilon$:
\begin{equation}
\nu = \frac{1}{2} + \frac{(N+2)}{4(N+8)}\,\epsilon + O(\epsilon^2)
\end{equation}

Finally, we comment briefly on the next-to-leading corrections at
$T=0$ to the Landau damping at the QCP. These are obtained from
corrections to scaling, i.e. depend on the manner in which $\gl$
flows to the fixed point:
\begin{equation}
\gl = g^* + (g_0-g^*) \lambda^\Omega + \cdots
\end{equation}
with:
\begin{equation}
\Omega = \beta'(g*) = \epsilon + O(\epsilon^2)
\end{equation}
The RG equation then yields the following form for the 2-point function, including
corrections to scaling:
\begin{equation}
\Gamma(q,\omega) = q^{2-\eta} \left[\gamma_0(\omega q^{-z})+
q^{2\Omega} \gamma_2(\omega q^{-z}) + \cdots\right]
\end{equation}
Taking the limit $q\rightarrow 0$, this yields the following form of the
Landau damping term:
\begin{equation}
\Gamma(q=0,i\omega) = C_0 |\omega|^{\frac{2-\eta}{z}} + C_2
|\omega|^{\frac{2-\eta+\Omega}{z}} + \cdots \label{subdominant}
\end{equation}
In the present case, the exact identity $z=2-\eta$ implies that the dominant term is
simply $\propto |\omega|$, while a correction of the form $|\omega|^{1+\frac{\Omega}{2-\eta}}$
is found. The logarithmic correction found above can be seen as the limiting behavior of this
correction in the marginal case $d=2$ ($\epsilon=0$).

\section{Large-N analysis of the $\phi^4$ theory of an antiferromagnetic
quantum critical point in two dimensions ($z=2$)}
\label{largeNlimit}

\begin{figure}[htbp]
\begin{center}
\vspace{0.6cm}
\includegraphics[width=9cm,angle=0]{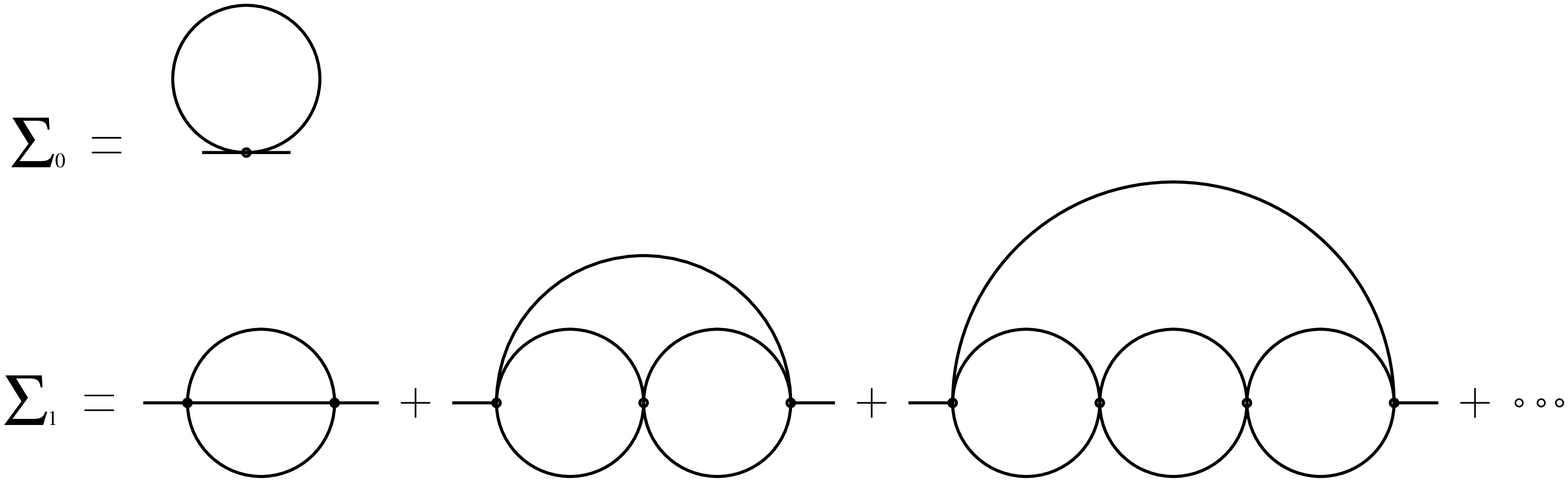}
\end{center}
\caption{Self-energy graphs in the $1/N$ expansion.}
\label{segraphs}
\end{figure}

In this section, we treat the field-theory (\ref{action}) within the large-N expansion.
For this purpose, we scale the coupling constant in (\ref{action}) by $1/N$ and set:
\begin{equation}
u_0 = \frac{6u}{N}
\label{scaled_u}
\end{equation}
We denote by $\Sigma_0$ and $\Sigma_1$ the contributions to the self-energy of
order $1/N^0$ and $1/N$, respectively
($\langle\phi\,\phi\rangle^{-1}=D_0^{-1}-\Sigma$).
The (skeleton) diagrams contributing to $\Sigma_0$ and $\Sigma_1$ are depicted in
Fig.~\ref{segraphs}. The corresponding expressions read:
\begin{equation}
\Sigma_0(q, i\omega)=-u T\sum_{\nu_n}\int \frac{d^2k}{(2\pi)^2}
D(k,i\nu_n) \label{sigma0}
\end{equation}
\begin{equation}
\Sigma_1(q, i\omega)=-2{\frac{u}{N}} T\sum_{\nu_n} \int
\frac{d^2k}{(2\pi)^2}\frac{D(k+q,i\nu_n+i\omega_n)}{1+u\Pi_0(k,i\nu_n)}
\label{sigma1}
\end{equation}
where:
\begin{equation}
\Pi_0(q, i\omega)=
T\sum_{\nu_n}\int {\frac{d^2p}{(2\pi)^2}} 
D(p,i\nu_n)\,D(p+q,i\nu_n+i\omega_n) \label{def_Pi0}
\end{equation}
In this expression, $D$ denotes the full propagator to order
$1/N^0$ (i.e including the self-energy $\Sigma_0$). Hence, Eq.~(\ref{sigma0}) should be viewed as
a self-consistent equation for $\Sigma_0$.
We first consider this equation at $\omega=q=0$, defining a
''gap'' $\Delta(T)=r-\Sigma(q=0,i\omega=0)$ related to the
correlation length $\xi$ by $\Delta=\xi^{-2}$. $\Delta$ vanishes
($\xi$ diverges) at the $T=0$ QCP. At the saddle point
($N=\infty$) level $\Delta(T)=\Sigma_0(T=0)-\Sigma_0(T)$. We show
in Appendix~\ref{gapeqappx} that, at this order, the
self-consistent equation for the gap reads (see also
Ref.~\onlinecite{sachdev2}):
\begin{equation}
\frac{(2\pi)^2}{u}\tilde\Delta=
{\frac{1}{2}}\ln{\frac{\tilde\Delta}{\tilde\Delta+\tilde\Lambda_{q^2}}}+
\ln{\frac{\Gamma(\tilde\Delta)}{\Gamma(\tilde\Delta+\tilde\Lambda_{q^2})}}+
\tilde\Lambda_{q^2}\ln{\tilde\Lambda_{q^2}}-\tilde\Lambda_{q^2}
\label{gapeq}
\end{equation}
where $\Lambda_{q^2}$ is a cutoff for $q^2$,
$\tilde\Lambda_{q^2}=\Lambda_{q^2}/(2\pi T)$ and
$\tilde\Delta=\Delta/(2\pi T)$. In the limit $\tilde\Delta\ll1$ and
$\tilde\Delta\ll\tilde\Lambda_{q^2}$ Eq.(\ref{gapeq}) reduces to:
\begin{equation}
\frac{(2\pi)^2}{u}\tilde\Delta=
-{\frac{1}{2}}\ln{\tilde\Delta}
-\tilde\Delta\ln{\tilde\Lambda}
\label{gapeqreduced}
\end{equation}
We dropped the index in $\Lambda_{q^2}$ as it is the only scale we use.
In the limit $u\ln{\tilde\Lambda}\gg1$ the approximate solution of
Eq.(\ref{gapeqreduced}) is\cite{sachdev2}:
\begin{equation}
\Delta=\pi\,T\,\frac{\ln(\ln{\Lambda/2\pi T})}{\ln{\Lambda/2\pi T}}
\end{equation}
Hence, $\Delta$ vanishes faster than $T$ as
$T\rightarrow 0$. The contributions of the marginally irrelevant coupling are smaller
than temperature (or energy) itself. As we shall now see, this also applies to the
$T=0$ contributions to the Landau damping at the critical point, which is the main
quantity of interest in this paper.

\begin{figure}[htbp]
\begin{center}
\vspace{0.6cm}
\includegraphics[width=9cm,angle=0]{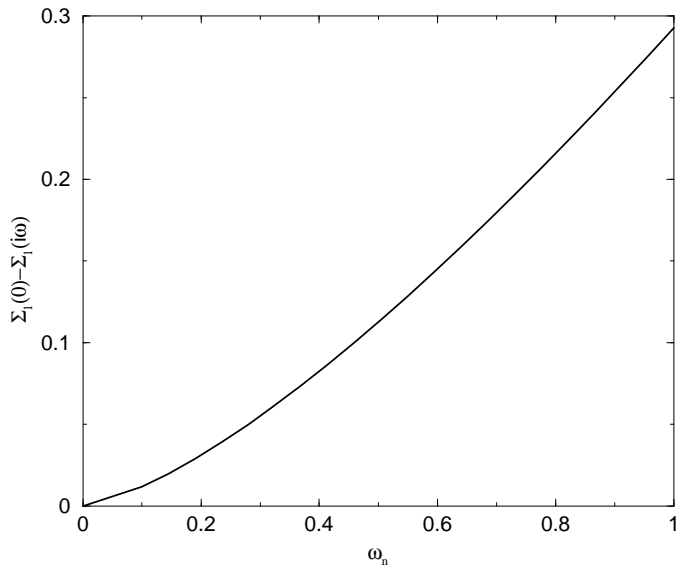}
\end{center}
\caption{Self-energy at the $d=2,z=2$ QCP, up to order $1/N$, as
a function of (imaginary) frequency $\omega$. The calculation is for
$u/(2\pi)^3=5$, and $\omega$ is measured
in units of the cutoff $\Lambda$.}
\label{dinv}
\end{figure}

\begin{figure}[htbp]
\begin{center}
\vspace{0.6cm}
\includegraphics[width=9cm,angle=0]{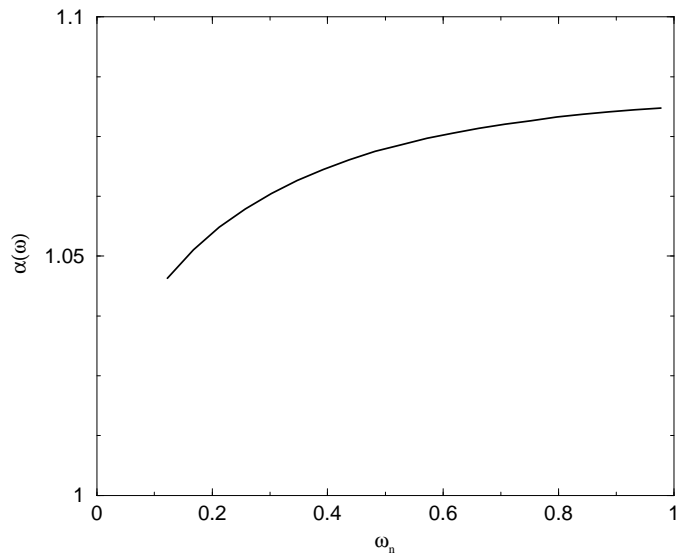}
\end{center}
\caption{Effective exponent $\alpha(\omega)$, as defined in the text, as
a function of (imaginary) frequency $\omega$.
The parameters are as in Fig.\ref{dinv}}
\label{alpha}
\end{figure}

We perform a calculation of the damping rate at the $T=0$ QCP up
to order $1/N$. At the saddle-point level ($1/N^0$), the
propagator at the QCP reads: $D(q, i\omega)=(|\omega|+q^2)^{-1}$.
The polarization bubble $\Pi_0(q, i\omega)$ defined by
(\ref{def_Pi0}) can then be calculated exactly. Details are given
in Appendix~\ref{polbubble}. Here we present the result:
\begin{eqnarray}
{8\pi^2}\Pi_0(q, i\omega)=&&\frac{\pi^2}{4}-
\frac{|\omega|+q^2}{q^2}\left(\ln{\frac{|\omega|+\Lambda}{\Lambda}}+
2\ln{\frac{|\omega|+q^2}{|\omega|}}\right)+
2\ln{\frac{|\omega|+\Lambda}{|\omega|}}
\nonumber\\
&&+\frac{s}{q^2}
\ln{\frac{|\omega|+q^2+2\Lambda+s}{|\omega|+q^2+2\Lambda-s}} -{\rm
Li}_2\left(\frac{q^2}{2|\omega|+q^2}\right) +{\rm
Li}_2\left(-\frac{q^2}{2|\omega|+q^2}\right)
\nonumber\\
&& +{\rm Li}_2\left(\frac{q^2}{|\omega|+s}\right) -{\rm
Li}_2\left(-\frac{q^2}{|\omega|+s}\right) +{\rm
Li}_2\left(\frac{q^2}{|\omega|-s}\right) -{\rm
Li}_2\left(-\frac{q^2}{|\omega|-s}\right) \label{chiwq}
\end{eqnarray}
where $s=\sqrt{(|\omega|+q^2)^2+4q^2\Lambda}$ and ${\rm
Li}_2(x)=-\int_0^x dy\ln(1-y)/y$. The above expression has the
following asymptotics at $\omega/\Lambda\to0$, $q^2/\Lambda\to0$:
\begin{eqnarray}
4-2\ln{\frac{q^2}{\Lambda}},\,\,\,\,\,\,\,\,\,\,\,\,\,\,
&\,\,\,\,\, \displaystyle\frac{|\omega|}{q^2} \rightarrow 0 &\nonumber\\
2+\frac{\pi^2}{4}-2\ln{\frac{|\omega|}{\Lambda}},
&\,\,\,\,\,\displaystyle\frac{q^2}{|\omega|}\to0,
&\,\,\,\,\,\frac{\omega^2}{q^2\Lambda}\to0\nonumber\\
-2+\frac{\pi^2}{4}-2\ln{\frac{|\omega|}{\Lambda}},
&\,\,\,\,\,\displaystyle\frac{q^2}{|\omega|}\to0,
&\,\,\,\,\,\frac{q^2\Lambda}{\omega^2}\to0
\end{eqnarray}
We use the explicit expression (\ref{chiwq}) into (\ref{sigma1})
and perform the frequency and momentum integrals numerically. The
resulting self-energy $\Sigma_1(q, i\omega)$ is plotted in
Fig.\ref{dinv}. It is seen that the $1/N$ corrections to the
damping rate are {\it less singular} than $\omega$ at
low-frequency. Alternatively, the corrections can be put in the
form of an effective, scale-dependent exponent $\alpha(\omega)$,
defined by:
$\alpha(\omega)=\omega\partial(\ln{\chi^{-1}(i\omega)})/\partial\omega$
with $\chi^{-1}(i\omega)=|\omega|-\Sigma_1(q,
i\omega=0)+\Sigma_1(q=0,i\omega=0)$ the $q=0$ dynamical
susceptibility at the QCP. This effective exponent is plotted on
Fig.~\ref{alpha}: it is seen to be only weakly dependent on
frequency in the range where it is displayed. Hence, the
logarithmic corrections in the marginal case can be mimicked by a
power-law, but correspond to an effective exponent $\alpha > 1$,
in contrast to the experimental observation $\alpha < 1$ for the
two compounds mentioned above.

\begin{figure}[htbp]
\begin{center}
\vspace{0.6cm}
\includegraphics[width=9cm,angle=0]{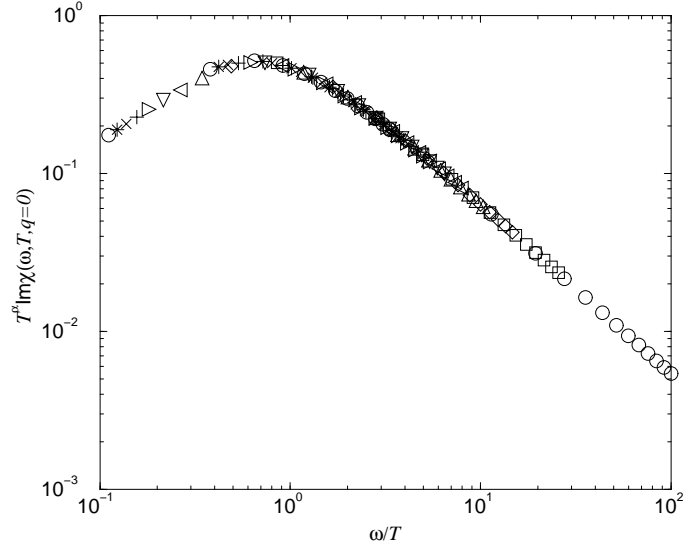}
\end{center}
\caption{Scaling plot for
$T^{\alpha}{\rm Im}\chi(\omega,T,q=0)$ versus
$\omega/T$. The temperatures displayed are:
$T=.03-.9\Lambda$.
The choice $\alpha=1.09$ provides a good collapse of the
data over this frequency range.
Different symbols correspond to different values of $T$.}
\label{collaps}
\end{figure}

\begin{figure}[htbp]
\begin{center}
\vspace{0.6cm}
\includegraphics[width=9cm,angle=0]{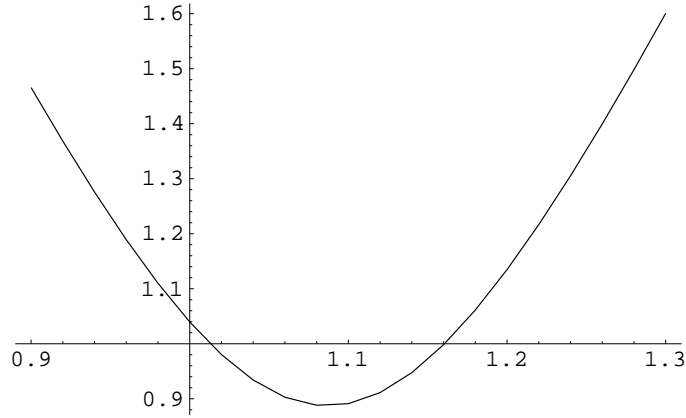}
\end{center}
\caption{Standard deviation $\sigma_{\log(T^{\alpha}\chi)}$, measuring the
quality of the $\omega/T$ collapse\cite{Schroder1998}, versus
$\alpha$.}
\label{error}
\end{figure}

Incidentally, we would like to comment on the data analysis made
in Ref.~\onlinecite{Schroder1998} in order to support $\omega/T$
scaling of the inelastic neutron scattering data in ${\rm
CeCu}_{5.9}{\rm Au}_{0.1}$. To this aim, we perform a similar
analysis on our analytical result, by plotting
$T^{\alpha}\chi(\omega,T)$ versus $\omega/T$ (Fig.~\ref{collaps}).
In this plot, the frequency-dependent susceptibility at finite
temperature and $q=0$ is approximated by:
$\chi(\omega,T)^{-1}\simeq\,
-i\omega+\Delta(T)-\Sigma_1(q=0,\omega)+\Sigma_1(q=0,\omega=0)$
(this approximation does not include the temperature dependence of
the Bose functions in the calculations which would be required to
obtain the correct damping coefficient, as is required to obtain
(\ref{noncommute})). Such a scaling plot is attempted for a
varying range of $\alpha$, and the optimal value of the exponent
is chosen such as to yield the best collapse\cite{Schroder1998},
as measured by the standard deviation plotted in Fig.~\ref{error}.
It is seen that an excellent collapse is obtained with
$\alpha\simeq 1.09$. We emphasize that $\omega/T$ scaling, in a
strict asymptotic sense, does apply here, but with a trivial
Gaussian exponent and scaling function:
$T\chi(\omega,T)=i/\tilde{\omega}$ ($\tilde{\omega}=\omega/T$).
The logarithmic corrections characteristic of the marginal case
apparently mimic non-trivial scaling properties over quite an
extended range of $\omega/T$. This should serve as a warning for
the interpretation of the experimental results.

We end this section by considering the {\it local approximation} to the
damped $\phi^4$ theory in the marginal case, up to order $1/N$. In a previous
paper, Motome and two of us\cite{pankov} have investigated the EDMFT approximation
to the critical behavior of various models, and shown that this local approximation is
quite satisfactory above the upper critical dimension (i.e here for $d+z>4$), when the
quartic coupling is irrelevant. In this approach, all skeleton graphs are taken to be local.
For arbitrary $z$ and $d=2$, the local propagator at the $T=0$ QCP reads:
\begin{equation}
D(i\omega)=\int \frac{qdq}{2\pi} \frac{1}{|\omega|^{\frac{2}{z}}+q^2}
=\frac{1}{4\pi}\ln\left(\frac{\Lambda+|\omega|^{\frac{2}{z}}}
{|\omega|^{\frac{2}{z}}}\right)
\label{propag}
\end{equation}
Performing the Fourier transform, this yields the asymptotic behavior at long
(imaginary) time:
\begin{equation}
D(\tau)=\int_{-\infty}^{+\infty}{\frac{d\omega}{2\pi}}
{\frac{1}{4\pi}}\ln\left(\frac{\Lambda+|\omega|^{\frac{2}{z}}}{|\omega|^{\frac{2}{z}}}\right)
e^{-i\omega \tau}\approx{\frac{1}{(2\pi)^2|\tau|}}\int_0^{+\infty}
dx\ln{x^{-{\frac{2}{z}}}}\cos{x}={\frac{1}{4\pi z|\tau|}}
\end{equation}
The $1/N$ correction to the spin-fluctuation self-energy is:
\begin{equation}
\Sigma_1(\tau) = \frac{2u^2}{N}\,D(\tau)^3 \approx {\frac{1}{N}}{\frac{2u^2}{(4\pi)^3 z^3 |\tau|^3}}
\label{sigmat}
\end{equation}
which corresponds to the following low-frequency behavior
(with $\Lambda_1$ a cutoff of order $1$):
\begin{equation}
\Sigma_1(i\omega)\approx {\frac{1}{N}}{\frac{2u^2}{(4\pi)^3z^3}}
(-{\frac{3}{2}}+\gamma-\ln{\Lambda_1}+\ln{|\omega|})\omega^2
\end{equation}
For real frequencies (\ref{sigmat}) corresponds to corrections to the Landau
damping of the form ${\rm Im}\Sigma_1(\omega+i0^+)\propto \omega^2 Sign(\omega)$. This
is much smaller (by a factor of order $\omega$, up to logarithms) than the corrections
obtained from the direct calculation in $d=2$ detailed above (Fig.~\ref{dinv}). Hence, we conclude that
the local (EDMFT) approximation is not very reliable at the upper critical dimension (marginal
case). As we shall see in the following section, a numerical solution of
the EDMFT equations for the Kondo lattice using an approximate ''impurity solver''
based on the large-N expansion leads to a
similar conclusion: only subdominant corrections to the Landau damping rate are generated
instead of the anomalous (power-law like) enhancement observed experimentally.

\section{''Locally'' critical point: self-consistent large-N solution of the EDMFT equations}
\label{fullN}

\subsection{EDMFT of the Kondo lattice and mapping onto a local spin-fluctuation model}

It has been recently argued\cite{SiNature,coleman_poland} that the understanding of non-Fermi liquid
behavior in heavy-fermion compounds close to a QCP requires a formalism in which electronic
degrees of freedom and spin fluctuations can be treated on the same footing, at least as a
starting point. This is a non-trivial task, which is however made easier in the
context of the extended dynamical mean-field theory (EDMFT)\cite{sismith,kaju,sengupta}
of the Kondo lattice model, which we consider here.

Let us consider the Kondo lattice Hamiltonian with an explicit exchange coupling between
localized spins, chosen to be Ising-like:
\begin{equation}
H\,=\, \sum_{k\sigma} \epsilon_{k} c^\dagger_{k\sigma}c_{k\sigma}
+ J_K \sum_{i\s\s'} \vec{S}_i. c^\dagger_{i\s} \vec{\tau}_{\s\s'} c_{i\s'}
+ \sum_{i<j} I_{ij}\,S^z_i S^z_j
\end{equation}
In this expression, $J_K$ is the (antiferromagnetic) Kondo coupling, and $\vec{\tau}$
stands for the Pauli matrices.
EDMFT maps this model onto a local impurity model with effective action:
\begin{eqnarray}
\label{Seff}\nonumber
S &=& - \inte \inte' \sum_\s \left[c^\dagger_\s(\tau) \mcal{G}_0^{-1}(\tau-\tau') c_\s(\tau')
+ S^z(\tau) \chi_0^{-1}(\tau-\tau') S^z(\tau')\right] \\
&+& \inte J_K \sum_{\s\s'} \vec{S}. c^\dagger_\s \vec{\tau}_{\s\s'} c_{\s'}
\end{eqnarray}
Both ${\cal G}_0$ and $\chi_0$ are effective ''Weiss fields'' which must be determined
in such a way that the following self-consistency conditions hold:
\begin{eqnarray}
G(i\wn) &\equiv& < c^\dagger_\s(i\wn) c_\s(i\wn)>_S\,
= \int \mr{d}\epsilon \frac{\rho_0(\epsilon)}{i\wn - \Sigma_c(i\wn) - \epsilon}
\label{scc} \\
\chi(i\nu_n) &\equiv& <S^z(i\nu_n) S^z(-i\nu_n)>_S\,
= \int \mr{d}\epsilon \frac{\rho_I(\epsilon)}{M(i\nu_n) - \epsilon}
\label{sci}
\end{eqnarray}
In this expression, $\Sigma_c(i\wn)$ and $M(i\nu_n)$ are respectively the fermionic
and local spin self-energies calculated within the impurity problem (\ref{Seff}), {\it i.e.}:
\begin{eqnarray}
G^{-1}(i\wn) &=& \mcal{G}_0^{-1}(i\wn) - \Sigma_c(i\wn)\\
\chi^{-1}(i\nu_n) &=& -\chi_0^{-1}(i\nu_n) + M(i\nu_n)
\end{eqnarray}
The densities of states $\rho_0$ and $\rho_I$ associated respectively with
the conduction band and to the spin interactions read:
$\rho_0(\epsilon)=\sum_k\delta(\epsilon-\epsilon_k)$,
$\rho_I(\epsilon)=\sum_q\delta(\epsilon-I_q)$.
The self-consistency conditions (\ref{scc},\ref{sci}) simply express that the impurity model Green's functions
and self-energies should coincide with their lattice counterpart (assuming momentum-independence
of the self-energies). They can be recast into a more compact form, relating directly
the local correlation functions to the two Weiss fields appearing in the effective
action (\ref{Seff}):
\begin{eqnarray}
\label{self1}
G(i\wn) &=& \int \mr{d}\epsilon \frac{\rho_0(\epsilon)}{i\wn
+ G^{-1}(i\wn) - \mcal{G}_0^{-1}(i\wn) - \epsilon}\\
\label{self2}
\chi(i\nu_n) &=& \int \mr{d}\epsilon
\frac{\rho_I(\epsilon)}{\chi^{-1}(i\nu_n) + \chi_0^{-1}(i\nu_n)  - \epsilon}
\end{eqnarray}
The key question at this stage is whether the electronic degrees of freedom can be
integrated out, particularly near the QCP. The local EDMFT framework allows to handle this issue
in a more controlled manner. Indeed, electronic degrees of freedom can be integrated
out\cite{sengupta} from the impurity model (\ref{Seff}) using an expansion in the spin-flip part of the Kondo
coupling \`a la Anderson-Yuval-Hammann\cite{anderson,chakra}.
This can be conveniently performed, for example, using bosonization
methods\cite{KotliarSi,grempel-roz,GrempelSi}. This leads to the
following action for the spin degrees of freedom:
\begin{equation}
\label{S_Ising}
S = \inte \inte' S^z(\tau) \left[ -\chi_0^{-1}(\tau-\tau')
+ K(\tau-\tau') \right] S^z(\tau')
+ \inte \; \Gamma S^x
\end{equation}
In this expression, $K(\tau-\tau')\sim 1/(\tau-\tau')^2$ is a long-range interaction
induced by the electronic modes. It corresponds to the bare Landau damping
($K(i\nu_n)=\kappa|\nu_n|$ at low-frequency).
The coupling in front of the spin-flip term is proportional to the Kondo
coupling: $\Gamma\propto \tau_0J_K$ (with $\tau_0$ a short-time cutoff).
Hence, the problem has been mapped onto an quantum Ising model in a transverse
field with long range interactions\cite{sengupta,grempel-roz,GrempelSi}.

\subsection{Numerical solution based on the large-N approximation}
\label{solN}

This mapping has been used in order to perform
analytical\cite{sengupta} and numerical\cite{grempel-roz} studies
of the behavior at the QCP in mean-field models with random
exchange couplings $I_{ij}$. It was demonstrated that this case is
formally similar\cite{footnote1} to the Hertz-Moriya-Millis theory
of an antiferromagnetic QCP in $d=3$. Recently, Grempel and Si
have analyzed the EDMFT theory of the QCP in the {\it
two-dimensional} case using this mapping\cite{GrempelSi}. Their
claim is that anomalous Landau damping
($\propto\omega^\alpha\,,\alpha<1$) and $\omega/T$ scaling can be
demonstrated in this case. Here, we present a solution of this
problem in which the local impurity problem is solved using a
large-N method. In contrast to the claims of
Ref.~\onlinecite{GrempelSi}, we find that only subdominant
corrections to the Landau damping are generated.

Following Ref.~\onlinecite{sengupta}, we deal with the local
action(\ref{S_Ising}) for the transverse field Ising model by
extending it to an $N$-component rotor model $\vec{n}$:
\begin{equation}
\label{rotor}
S = \frac{1}{2}\inte \; \left[ \frac{(\dt \vec{n})^2}{g} + \lambda(\tau) (\vec{n}^2 - N) \right]
+ \frac{1}{2} \inte \inte' \; \vec{n}(\tau) \left[ -\chi_0^{-1}(\tau-\tau')
+ K(\tau-\tau') \right] \vec{n}(\tau') \\
\end{equation}
In this expression, $g$ plays a role to similar to that of the transverse field
in the original Ising model (disordering term).
We define a self-energy correction for the local spin susceptibility
$\chi(\tau)\equiv\langle S^z(0)S^z(\tau)\rangle$:
\begin{equation}
\label{sigma_def}
\chi(i\nu_n) = \frac{1}{ \nu_n^2/g + K(i\nu_n)
- \chi_0^{-1}(i\nu_n) - \Sigma(i\nu_n) }
\end{equation}
Let us first consider the problem at the saddle-point level ($N=\infty$), for a given
Weiss function $\chi_0$. The self-energy is then frequency- independent
(as noted by Grempel and Si\cite{GrempelSi})
and is given by
the saddle-point value of the Lagrange multiplier:
\begin{equation}
\Sigma_0(i\nu_n) = -\lambda
\label{sigma_saddle}
\end{equation}
such that:
\begin{equation}
\chi_\infty(\tau=0) = 1
\label{eq_saddle}
\end{equation}
In this expression, $\chi_\infty$ is the local susceptibility at the
saddle-point ($N=\infty$) level:
\begin{equation}
\chi_{\infty}(i\nu_n) = \frac{1}{ \nu_n^2/g + K(i\nu_n)
- \chi_0^{-1}(i\nu_n)+\lambda}
\label{chi_saddle}
\end{equation}
Both $\lambda$ and $\chi_{\infty}$ depend on the Weiss function $\chi_0$.
We approximate the self-energy by its first two terms in the $1/N$ expansion:
\begin{eqnarray}
\label{sigma_sub}
\Sigma(i\nu_n) &\simeq& \Sigma_0(i\nu_n) + \Sigma_1(i\nu_n) - \Sigma_1(i0)\\
\label{sigma_subb}
\Sigma_1(\tau) &=&  -\frac{2}{N} \Gamma(\tau) \chi_\infty(\tau)\\
\label{sigma_subbb}
\mr{with} \;\;\; \Gamma(i\nu_n) &=& \frac{1}{\Pi(i\nu_n)} \;\;\;
\mr{and} \;\;\; \Pi(\tau) = \left[ \chi_\infty(\tau)\right]^2
\end{eqnarray}
The EDMFT equations are thus solved by iterating (numerically) the following procedure.
For a given $\chi_0(i\nu_n)$, the saddle-point quantities
$\lambda=-\Sigma_0$ and $\chi_{\infty}$ are computed by solving
Eqs.~(\ref{eq_saddle},\ref{chi_saddle}). Then, the self-energy is computed from
(\ref{sigma_sub},\ref{sigma_subb},\ref{sigma_subbb}). This is inserted into the
EDMFT self-consistency condition in order to obtain the local susceptibility:
\begin{equation}
\chi(i\nu_n) = \int \mr{d}\epsilon
\frac{\rho_I(\epsilon)}{\nu_n^2/g + K(i\nu_n)-\Sigma(i\nu_n)-\epsilon}
\end{equation}
From this, an updated value of the Weiss function is obtained as:
$\chi_0^{-1}(i\nu_n)=\nu_n^2/g +
K(i\nu_n)-\Sigma(i\nu_n)-\chi^{-1}(i\nu_n)$, and the procedure is
iterated until convergence is reached.

\begin{figure}[htbp]
\begin{center}
\vspace{0.6cm}
\includegraphics[width=10cm]{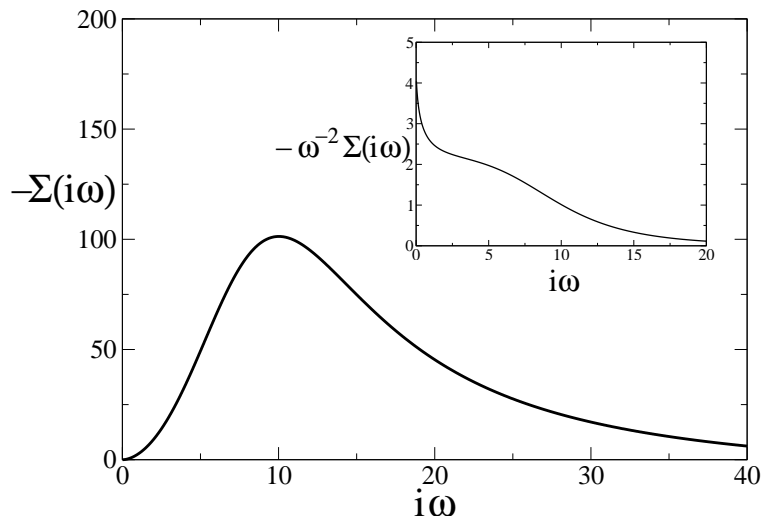}
\end{center}
\caption{Self-energy $\Sigma(i\w)$ close to the critical point (at inverse
temperature $\beta=500$ and $g=g_c=0.223$), as a function of
imaginary frequency. For this calculation,
we have set $N=2$, and used a flat density of states
$\rho_I(\epsilon)$ of half-width equal to unity.
Inset: plot of $\w^{-2}\Sigma(i\w)$, showing the logarithmic behavior of
this quantity at small frequencies.}
\label{sigma_edmft}
\end{figure}

\begin{figure}[htbp]
\begin{center}
\vspace{1.0cm}
\includegraphics[width=10cm]{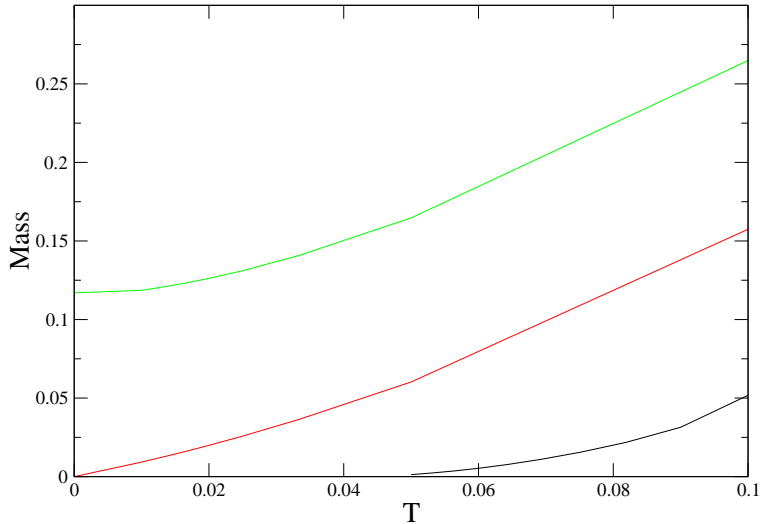}
\end{center}
\caption{Plot of the mass $\Sigma(i0)$ as a function of temperature,
demonstrating the existence of a continuous transition at $T=0$ within our
approximate solution of the EDMFT equations. From bottom to top:
$g = 0.1$ (ordered state), $g=0.223$ (critical at $T=0$),
$g=0.3$ (disordered case).}
\label{mass}
\end{figure}

The result of this numerical calculation for the (local) spin self-energy is
displayed in Fig.~\ref{sigma_edmft}. The behavior
of the self-energy at low imaginary frequency is well fitted by
$\Sigma_1(i\omega) \sim \w^2 \ln|\w|$ (inset) when one tunes the parameters to sit at
criticality (Fig. \ref{mass}).
This corresponds to subdominant corrections to the Landau damping
${\rm Im}\Sigma(\omega+i0^+)\propto\omega^2 {\rm sign}(\omega)$.
These findings are in agreement with the expansion of the EDMFT equations
for the $\phi^4$ model at order $1/N$ discussed at the end of the previous section,
and provide an independent check of this result. The method followed in the present section
has been to use the large-N method as an approximate ''impurity solver'' for the
local problem. It can be easily checked that if, instead, the equations used here are
all expanded up to order $1/N$, the equations of the local approximation
presented in the previous section are recovered.


\section{Conclusion}

We analyzed two of the  approaches to  a critical theory
of the antiferromagnetic metal to paramagnetic metal
in the ${\rm CeCu}_{6-x}{\rm Au}_x$
and ${\rm YbRh}_2{\rm Si}_2$ materials from the perspective
of the antiferromagnetic spin fluctuations.

In the context of the model of Rosch et. al.\cite{Rosch}
we analyzed the corrections to mean field theory due to
the marginally  irrelevant fourth order coupling.
We find that  while  there are logarithmic corrections
to mean field theory which vanish as $1/{\rm Log}^2|\omega|$
the coefficient of this corrections is such that it can
only  be mimicked by a decreased Landau damping, in
clear disagreement
with experimental observations.
In our view, this rules out this model, in spite
of the fact that it can account nicely for many of the
electronic properties of this systems.

We also analyze  the  "local " version of this model
using the large N expansion. Again in this case, we
find that it is not possible to account  for the anomalous frequency
dependence  of the damping coefficient.
This is in disagreement with reference \cite{GrempelSi}
but in agreement  with a recent EDMFT study of the Anderson  lattice
model.\cite{PingSun}

The anomalous damping has been recently observed in NMR
measurements in both\cite{ishida,walstedt} ${\rm YbRh}_2{\rm
Si}_2$ and ${\rm CeCu}_{6-x}{\rm Au}_x$. While it  has different
temperature dependencies in each  system, it is clearly enhanced
above the fermi liquid temperature dependence. Possible resolution
of this puzzle may require a better treatment of the effects of
disorder
or better treatments of models which retain the fermionic nature
of the spin fluctuations as proposed by Coleman et. al.\cite{coleman_poland}
and Si et. al.\cite{SiNature}

\begin{acknowledgments}
We thank D. Grempel and Q. Si for useful discussions and correspondence.
The authors acknowledge the hospitality of the Kavli Institute for
Theoretical Physics (UCSB, Santa Barbara), where part of this work was
performed and supported under grant NSF-PHY99-07949 and
NSF grant No. DMR-0096462. We also acknowledge the
support of an international collaborative grant from CNRS (PICS 1062).
\end{acknowledgments}

\appendix

\section{Computation of RG flow}
\label{rgflowappx}

The computations leading to the RG flow (\ref{beta_eta}) are
standard. Here we only provide some further details of the
computation of the number $c_2$, and of the constant $c$ required
to obtain the universal logarithmic correction in
(\ref{rg_correction}). These computations involve the $|\omega|$
term in the propagator in an essential and novel manner.

Repeating the standard field-theoretic derivation for the two-loop
graph in Fig~\ref{I1I2} we have:
\begin{eqnarray}
I_2 (q,i\omega=0) &=& \frac{u_0^2 (N+2)}{18}
\int_{-\infty}^{\infty} \frac{d \omega_1}{2 \pi} \frac{d
\omega_2}{2 \pi} \int \frac{d^d q_1}{(2\pi)^d} \frac{d^d
q_2}{(2\pi)^d} \frac{1}{(q_1^2 + |\omega_1|)(q_2^2 +
|\omega_2|)} \nonumber \\
&~&~~~~~~~~~~~~~\times \frac{1}{(({\bf q} + {\bf q_1} + {\bf
q_2})^2 + |\omega_1+\omega_2|)} = g^2 q^{2d-2} \left( -
\frac{c_2}{2\epsilon} + \mathcal{O} (\epsilon^0) \right)
\label{s2}
\end{eqnarray}
$I_2$ is to be evaluated for $d<1$, where it is convergent, and
the result analytically continued to near $d=2$, where it is
expected have the small $\epsilon$ expansion shown on the r.h.s.
The coefficient of the pole fixes the value of the constant $c_2$.

Note that we are evaluating $I_2$ in an external momentum ${\bf
q}$, and in $d=2$ the pole on the r.h.s. corresponds to the
appearance of $q^2 \ln q$ term. In contrast, evaluation of $I_2
(0,i\omega)$ (presented below) shows that no pole appears, and
hence there is no $|\omega| \ln |\omega|$ term in $d=2$. This
absence of such a pole is, of course, the reason for the absence
of an independent renormalization constant for the damping term
noted below (\ref{def:zl}), and for the $\omega$ dependence of
$I_2$ noted above (\ref{rg_correction}). It is also responsible
for the exponent identity $z=2-\eta$ for $d<2$.

We now turn to the evaluation of $I_2$ in (\ref{s2}) in an
external momentum ${\bf q}$. First, we evaluate the integral over
$q_{2}$ by the standard Feynman parameter method
\begin{eqnarray}
I_2 (q,0) &=& \frac{u_0^2 (N+2)\Gamma (2-d/2)}{18(4 \pi)^{d/2}}
\int_{-\infty}^{\infty} \frac{d \omega_1}{2 \pi} \frac{d
\omega_2}{2 \pi} \int \frac{d^d q_1}{(2\pi)^d} \int_0^1 dx
\frac{1}{(q_1^2 + |\omega_1|)} \nonumber \\
&~&~~~~~~\times\frac{1}{\left[({\bf q} + {\bf q}_1)^2 x(1-x) +
|\omega_2|(1-x) + |\omega_1 + \omega_2| x\right]^{2-d/2}}.
\label{s3}
\end{eqnarray}
Similarly, performing the integral over $q_1$ with a Feynman
parameter $y$ we obtain
\begin{eqnarray}
I_2 (q,0) &=& \frac{u_0^2 (N+2)\Gamma(3-d)}{18(4 \pi)^d}
\int_{-\infty}^{\infty} \frac{d \omega_1}{2 \pi} \frac{d
\omega_2}{2 \pi} \int_0^1 dx
[x(1-x)]^{\epsilon/2} \int_0^1 dy y^{\epsilon/2} \nonumber \\
&~&\times \frac{1}{[q^2 x(1-x)y(1-y) + |\omega_1 + \omega_2|
x(1-x)(1-y) + |\omega_1| y(1-x) +  |\omega_2| xy]^{1+ \epsilon}}
\label{s4}
\end{eqnarray}
Now we perform the integral over $\omega_1$ and $\omega_2$ by
using the useful formula
\begin{equation}
\int_{-\infty}^{\infty} d \omega_1 d \omega_2 \frac{1}{[A + B
|\omega_1| + C |\omega_2| + D |\omega_1 + \omega_2|]^{\sigma}} =
\frac{4 A^{2-\sigma} (B+C+D) \Gamma(\sigma-2)}{(B+C)(C+D)(B+D)
\Gamma(\sigma)}. \label{s5}
\end{equation}
The formula (\ref{s5}) is derived by explicitly performing the
integrals over $\omega_{1,2}$ over different regions in the
$\omega_{1,2}$ plane delineated by changes in signs of $\omega_1$,
$\omega_2$, and $\omega_1+\omega_2$. Note that (\ref{s5}) is to be
evaluated at $\sigma = 1 + \epsilon$. Consequently, we obtain from
(\ref{s5}) the factor $\Gamma(\sigma-2) =
\Gamma(-1+\epsilon)=-1/\epsilon  +\mathcal{O} (\epsilon^0 )$,
which gives us the requisite pole in $\epsilon$ appearing on the
r.h.s. of (\ref{s2}). So we may safely set $d=2$ in the remaining
terms in (\ref{s4}). In this manner, combining (\ref{s2}),
(\ref{s4}) and (\ref{s5}) we obtain
\begin{equation}
c_2=\frac{(N+2)}{36} \int_0^1 dx \int_0^1 dy
\frac{(1-y)\left[x(1-x)(1-y) + y(1-x) + xy\right]}{[x(1-y) +
y][(1-x)(1-y)+y]}
=\frac{(N+2)(12-\pi^2)}{144}. \label{s6}
\end{equation}

Let us now present a few details of the evaluation of $I_2
(0,i\omega)$. In this case, (\ref{s4}) is replaced by
\begin{eqnarray}
I_2 (0,i\omega) &=& \frac{u_0^2 (N+2)\Gamma(3-d)}{18(4 \pi)^d}
\int_{-\infty}^{\infty} \frac{d \omega_1}{2 \pi} \frac{d
\omega_2}{2 \pi} \int_0^1 dx
[x(1-x)]^{\epsilon/2} \int_0^1 dy y^{\epsilon/2} \nonumber \\
&~&\times \frac{1}{[ |\omega + \omega_1 + \omega_2| x(1-x)(1-y) +
|\omega_1| y(1-x) +  |\omega_2| xy]^{1+ \epsilon}}. \label{s7}
\end{eqnarray}
Now, instead of (\ref{s5}), we need the following integral
\begin{eqnarray}
&& \int_{-\infty}^{\infty} d \omega_1 d \omega_2 \frac{1}{[A
|\omega_1| + B |\omega_2| + C |\omega + \omega_1 +
\omega_2|]^{\sigma}} = \frac{4 |\omega|^{2-\sigma} ABC
\Gamma(\sigma-2)}{\Gamma(\sigma)} \nonumber \\
&~&~~~~~\times \left[ \frac{A^{1-\sigma}}{(A^2-B^2)(A^2-C^2)} +
\frac{B^{1-\sigma}}{(B^2-A^2)(B^2-C^2)} +
\frac{C^{1-\sigma}}{(C^2-A^2)(C^2-B^2)}\right] . \label{s8}
\end{eqnarray}
As below (\ref{s5}), we need to evaluate (\ref{s8}) at $\sigma = 1
+ \epsilon$ and pick out a possible pole in $\epsilon$. Indeed, a
possible pole does appear to be present in the $\Gamma(\sigma-2)$
pre-factor in (\ref{s8}). However, careful evaluation shows that
that the residue of such a pole vanishes, and (\ref{s8}) in fact
has a smooth $\sigma \rightarrow 1$ limit:
\begin{eqnarray}
\lim_{\sigma \rightarrow 1} (\mbox{\protect\ref{s8}}) = 4 ABC
|\omega| \left[ \frac{\ln A}{(A^2-B^2)(A^2-C^2)} + \frac{ \ln
B}{(B^2-A^2)(B^2-C^2)} + \frac{\ln C}{(C^2-A^2)(C^2-B^2)}\right]
\label{s9}
\end{eqnarray}
The $\epsilon \rightarrow 0$ limit of the remaining terms in
(\ref{s7}) is straightforward, and this establishes the claimed
absence of a $|\omega|\ln|\omega|$ term in
$I_2 (0,i\omega)$ in $d=2$. Evaluating the constant $c$ in
$I_2 (0,i\omega)=I_2(0,0)-cg^2|\omega|$ we obtain:
\begin{equation}
c=\frac{(N+2)}{36}\,\, \mathcal{P}\int_0^1 dx \int_0^1 dy
\frac{x^2(1-y)}{(1-2x)\left[y^2-x^2(1-y)^2\right]}
\ln{\left(x\frac{1-y}{y}\right)}=
\frac{(N+2)(6\pi^2\ln{2}-11\zeta(3))}{1728}
\label{c}
\end{equation}
We performed integrations in Eq.(\ref{c}) by changing variables
to $x'=x$, $y'=x(1-y)/y$ and integrating first over $x'$ and
then over $y'$.

\section{Gap equation}
\label{gapeqappx}

In this appendix we derive the gap equation Eq.(\ref{gapeq}).
We need to evaluate:
\begin{equation}
\Sigma_0(T)=-uT\sum_{\omega_n}\int \frac{qdq}{2\pi} D(q,i\omega)=
{\frac{1}{4\pi}}uT\sum_{\omega_n}\ln{\frac{|\omega_n|+\Delta}
{|\omega_n|+\Delta+\Lambda_{q^2}}}
\end{equation}
We consider a $q^2$ cutoff $\Lambda_q^2$, which from now on we denote
as $\Lambda$, and a frequency cutoff $\Lambda_{\omega}\in 2\pi n T$
which is to be taken to infinity. For this choice of the sharp cutoff
$\Lambda_{\omega}$ one can show:
\begin{equation}
S_1(T)=\sum_{-\Lambda_{\omega}<\omega_n<\Lambda_{\omega}}
\ln{\frac{|\omega_n|+\Delta}{|\omega_n|+\Delta+\Lambda_{q^2}}}=
\ln{\frac{\tilde\Delta+\tilde\Lambda}{\tilde\Delta}}
+2\ln{{\frac{\Gamma(\tilde\Delta+\tilde\Lambda)
\Gamma(\tilde\Delta+\tilde\Lambda_{\omega})}
{\Gamma(\tilde\Delta)
\Gamma(\tilde\Delta+\tilde\Lambda+\tilde\Lambda_{\omega})}}}
\label{s1}
\end{equation}
where $\tilde\Lambda=\Lambda/(2\pi T)$,
$\tilde\Lambda_{\omega}=\Lambda_{\omega}/(2\pi T)$ and
$\tilde\Delta=\Delta/(2\pi T)$.
Taking the limit $\tilde\Lambda_{\omega}\to\infty$ we have:
\begin{equation}
S_1(T)=
\ln{\frac{\tilde\Delta+\tilde\Lambda}{\tilde\Delta}}
+2\ln{\frac{\Gamma(\tilde\Delta+\tilde\Lambda)}
{\Gamma(\tilde\Delta)}}-2\tilde\Lambda\ln{\tilde\Lambda_{\omega}}
\label{s1limit}
\end{equation}
The finite $T$ self energy is $\Sigma_0(T)={\frac{1}{4\pi}}uTS_1(T)$, and
the zero temperature $\Sigma_0(T=0)$ is:
\begin{equation}
\Sigma_0(T=0)=\left.{\frac{1}{4\pi}}uTS_1(T)\right|_{T\to0}=
\frac{u\Lambda}{(2\pi)^2}(\ln{\frac{\Lambda}{\Lambda_{\omega}}}-1)
\end{equation}
Noticing that $\Delta(T)=\Sigma_0(T=0)-\Sigma_0(T)$ we write
Eq.(\ref{gapeq}).

\section{Polarization bubble $\Pi_0$}
\label{polbubble}

Here we show in detail how we compute $\Pi_0(q, i\omega)$ in $d=2$
with $z=2$. We use a finite cutoff $\Lambda_{q^2}\equiv\Lambda$
for $q^2$ in a momentum integration, and we use
$\Lambda_{\omega}\to\infty$ as a frequency cutoff. We use the
Feynman parameterization to find:
\begin{eqnarray}
\Pi_0(q, i\omega)=\int {\frac{d^2p}{(2\pi)^2}} {\frac{d\nu}{2\pi}}
D(p,i \nu) D(p+q,i\nu+i\omega)&&
\nonumber\\
={\frac{1}{4\pi}}\int_0^1 dx \int_{-\infty}^{+\infty}{\frac{d\nu}{2\pi}} &&
\left(l(x,q,\nu,i\omega)^{-1}-[\Lambda+l(x,q,\nu,i\omega)]^{-1}\right)
\label{intq}
\end{eqnarray}
where $l(x,q,\nu,i\omega)=x(1-x)q^2+x|\nu|+(1-x)|\nu+\omega|$
and $D^{-1}=|\omega|+q^2$ at the QCP.
Integrating out frequency we get:
\begin{eqnarray}
\Pi_0(q, i\omega)={\frac{1}{8\pi^2}}\,\,\mathcal{P}\int_0^1 dx
\frac{4x}{1-2x}&&\left\{\ln{\left[(1-x)(q^2x+|\omega|)\right]}\right.
\nonumber\\
&&\left.-\ln{\left[\left({\frac{1}{2q^2}}(q^2-|\omega|+s)-x\right)
\left(q^2x+{\frac{1}{2}}(|\omega|-q^2+s)\right)\right]}\right\}
\label{afterfreq}
\end{eqnarray}
where $s=\sqrt{(|\omega|+q^2)^2+4q^2\Lambda}$. It is easy to
show:
\begin{equation}
\mathcal{P}\int_0^1 dx
\frac{4x}{1-2x}\ln(ax+b))=2-2\frac{a+b}{a}\ln{\frac{a+b}{b}}
-2\ln{b}-{\rm Li}_2(\frac{a}{a+2b})+{\rm Li}_2(-\frac{a}{a+2b})
\label{integral}
\end{equation}
for $a+b>0$ and $b>0$. From Eq.(\ref{afterfreq}) and
Eq.(\ref{integral}), after simple algebra, Eq.(\ref{chiwq})
follows.


\begin{references}
\bibitem{footnote1} It is not yet clear at this stage whether the EDMFT approach to the
antiferromagnetic Kondo lattice in the non-random case
follows the conventional Hertz-Millis behaviour, or undergoes a first-order transition
from the Fermi liquid to the ordered phase. A Large-N study for an $SU(N)$-symmetric
exchange coupling (in contrast to the Ising case studied here)
provides indications in favor of the latter possibility\cite{burdin}.
\end{references}
\end{document}